\documentclass[%
 aip,
 jmp,%
 amsmath,amssymb,
 reprint,%
]{revtex4-1}
\usepackage[colorlinks=true,linkcolor=blue]{hyperref}%
\usepackage{array,longtable}
\usepackage{graphics}
\usepackage{graphicx}
\usepackage{epsfig}

\def \be {\begin{equation}}
\def \ee {\end{equation}}
\def \ba {\begin{eqnarray}}
\def \ea {\end{eqnarray}}
\def \bm {\begin{displaymath}}
\def \em {\end{displaymath}}
\def \vr {{\roarrow r}}

\begin{document}
\title{ Integral Equation Theory for Pair Correlation 
Functions in a Crystal}
\author{Anubha Jaiswal, Atul S. Bharadwaj and Yashwant Singh}
\affiliation{Department of Physics, Banaras Hindu University, 
Varanasi-221 005, India 
\\
 }
\date{\today}
\begin{abstract}
A method for calculating pair correlation functions in a crystal is developed. The method is based 
on separating the one- and two- particle correlation functions into the symmetry conserving and the symmetry 
broken parts. The conserving parts are calculated using the integral equation theory of homogeneous 
fluids. The symmetry broken part of the direct pair correlation function is calculated from a series 
written in powers of order parameters and that of the total pair correlation function from the 
Ornstein- Zernike equation. The results found for a two-dimensional hexagonal lattice show that 
the method provides accurate and detailed informations about the pair correlation functions in a crystal. 
\end{abstract}
\pacs{61.50.Ah, 63.20.dk, 05.20.-y, 64.70.D-}
\maketitle

\section*{ }
The structural and thermodynamic properties of a classical system can 
adequately be described in terms of one and two-particle density distributions \cite{1,2}.
The one particle density distribution $ \rho ( \vec{r_{1}})$ gives the probability 
of finding a particle at position $ \vec{r_{1}} $ while the two-particle density 
distribution $ \rho^{(2)} (\vec{r_{1}},\vec{r_{2}}) $ gives the probability of finding 
simultaneously a particle at position $ \vec{r}_{1} $ and another particle at 
position $ \vec{r}_{2} $. The pair distribution function 
$ g(\vec{r}_{1},\vec{r}_{2}) $  which in the case of a homogeneous system 
reduces to the radial distribution function (RDF) 
$ g^{(0)} (|\vec{r}_{2} - \vec{r}_{1}|) $  is related to 
$\rho^{(2)} (\vec{r}_{1},\vec{r}_{2})$ through the relation 
$ \rho^{(2)} (\vec{r}_{1},\vec{r}_{2})  =  \rho(\vec{r}_{1}) \rho (\vec{r}_{2}) g(\vec{r}_{1},\vec{r}_{2})$.
The two related but different pair correlation functions (PCFs) that appear in the 
statistical theory of classical systems are the total pair correlation function 
$ h(\vec{r}_{1},\vec{r}_{2}) $ defined as $ h(\vec{r}_{1},\vec{r}_{2}) =  g(\vec{r}_{1},\vec{r}_{2}) -1 $ 
and the direct pair correlation function (DPCF) $ c(\vec{r}_{1},\vec{r}_{2}) $. 
These functions are related through the Ornstein-Zernike (OZ) equation.
In a homogeneous system, functions $ \rho^{(2)} $, h and c depend only on the
inter-particle separation and are function of density $ \rho =\frac{N}{V} $ (N being the number 
of particles in volume V), whereas in an inhomogeneous system 
they depend on position vectors of particles and are functional of $ \rho (\vec{r} ) $.

     A crystal is a system of extreme inhomogeneities where values of $ \rho (\vec{r}) $ shows 
several order of magnitude difference between its values on the lattice sites and in the 
interstitial regions. The periodic structure of crystals allows one to 
expand $\rho(\vr)$ in a Fourier series where the sum is carried over the reciprocal lattice vectors 
or to write it as a superposition of normalised Gaussians centred around the lattice 
points. The two-particle density distribution has been approximated  as 
$ \rho^{(2)} (\vec{r}_{1},\vec{r}_{2})  =  \rho(\vec{r}_{1}) \rho (\vec{r}_{2}) g(|\vec{r}_{2}-\vec{r}_{1}|,\bar{\rho})$  where $ g(|\vec{r}_{2}-\vec{r}_{1}|,\bar{\rho}) $ is the RDF  of a fluid 
of density $ \bar{\rho} $ which is taken to be much lower than the averaged crystal  
density $ \rho $ [3-6]. This amounts to assuming that 
 $ g(\vec{r}_{1},\vec{r}_{2}) $ is a function of magnitude of the 
interparticle separation $r = |\vr_{2} -\vr_{1}|$ only and also 
its value is much lower compared to that of a fluid of density $\rho$. 
Moreover, as the value of $g(r,\bar{\rho})$ decays exponentially, the 
assumption means that $ \rho (\vec{r}_{1}) 
\rho(\vec{r}_{2}) h(\vec{r}_{1},\vec{r}_{2}) $ is a short range function.
On the other hand, calculation of  $ \rho (\vec{r}_{1}) 
\rho(\vec{r}_{2}) h(\vec{r}_{1},\vec{r}_{2}) $
done in the harmonic model of crystals shows that this quantity  decays as 
$ |\vec{r}_{2} - \vec{r}_{1}|^{-1} $ in three-dimensions 
\cite{7,8}. Obviously, there is a need to have a theory for PCFs in crystals   
analogous to the integral equation theory (IET) of homogeneous fluids.
Accurate knowledge of PCFs will provide a unified approach to describe 
both states of matter including the 
freezing/melting and the solid-solid transitions.

 The IET which is used to calculate PCFs in a system 
interacting via a known pair potential, consists of the OZ equation and a closure relation 
that relates PCFs to the pair potential \cite{2}. The theory has been used 
successfully to find values of PCFs h and c in fluids from 
zero density to close to the freezing density. The application of the theory 
to crystals or to other inhomogeneous systems has, however, so far been limited
\cite{9}. In this context, it is important to note that while the OZ equation 
is general and applicable to fluids as well as to crystals, the closure relations 
which have been derived assuming continuous translational symmetry are valid only in 
homogeneous fluids but not in crystals. This limits the applicability of the existing 
IET to crystals.

  The method proposed here uses the OZ equation and some relations (Eqs (6)-(8)) of
 density functional theory \cite{2}. The method has following steps:
(I) One- and two- particle distribution functions are written as 
a sum of two contributions; one which preserves the continuous symmetry 
of the fluid and the other that arises due to breaking of this symmetry 
at the freezing point. 
(II) Separation of the OZ equation into two equations; one which contains 
the symmetry conserving part of $ \rho (\vec{r}) $,  $ h(\vec{r}_{1},\vec{r}_{2}) $ and 
$ c(\vec{r}_{1},\vec{r}_{2}) $ and the other in which the symmetry broken part of 
these functions appear. 
(III) Evaluation of symmetry conserving part of correlations and their derivatives 
with respect to density $ \rho $ using the IET of homogeneous systems.
(IV) Evaluation of the symmetry broken part of $ c(\vec{r}_{1},\vec{r}_{2}) $ 
using a series expressed in powers of order parameters (order parameters are 
amplitudes of density waves of wavelength $ \frac{2\pi}{|G|} $ ; $ \vec{G} $ being the reciprocal lattice 
vectors (RLVs.). 
(V) Using known values of $ c(\vec{r}_{1},\vec{r}_{2}) $ the OZ equation is solved to find
$ h(\vec{r}_{1},\vec{r}_{2}) $ for a given $ \rho (\vec{r}) $. We describe these steps below and
report results for a two-dimensional solid of hexagonal symmetry and use Ward identities to 
check their accuracy. 

  The OZ equation that connects $ h $ and $ c $ functions 
in an inhomogeneous system is, 
\begin{equation}
h(\vec{r}_{1},\vec{r}_{2}) = c(\vec{r}_{1},\vec{r}_{2}) + \int d \vec{r}_{3} c(\vec{r}_{1},\vec{r}_{3}) \rho (\vec{r}_{3}) h(\vec{r}_{2},\vec{r}_{3}) ,   \label{1}
\end{equation}
 The $ \rho(\vec{r}) $ in a crystal can be written as a sum of two terms 
 $ \rho(\vec{r}) = \rho + \rho^{b} (\vec{r}) $ where 
$ \rho^{b} (\vec{r}) = \sum_{G} \rho_G \exp (i\vec{G} \cdot \vec{r}) $
 . Here $ \rho $ is the average density of the crystal, $ \rho_{G} $ are the order 
 parameters and sum is over the complete set of RLVs. We refer the first term $\rho$ 
 as symmetry conserved and the second $\rho^{(b)}(\vr) $ as symmetry broken parts of
 $ \rho(\vec{r}) $. Similarly PCFs $ h $ and $ c $ can be written as a sum of 
 symmetry conserved and symmetry broken parts \cite{10};
\begin{eqnarray}
h(\vec{r}_{1},\vec{r}_{2})=h^{(0)}(|\vec{r}_{2}-\vec{r}_{1}|,\rho) + h^{(b)}(\vec{r}_{1},\vec{r}_{2},[\rho]) \nonumber 
\\
c(\vec{r}_{1},\vec{r}_{2})=c^{(0)}(|\vec{r}_{2} - \vec{r}_{1}|,\rho) + c^{(b)}(\vec{r}_{1},\vec{r}_{2},[\rho]).
 \label{2}
\end{eqnarray}
The symmetry conserved part $(h^{(0)}, c^{(0)})$ depends on magnitude of the 
interparticle separation  $r = |\vec{r}_{2} - \vec{r}_{1}|$ and is a function of density $\rho $,
whereas the symmetry broken part $(h^{(b)}, c^{(b)})$ is functional of $\rho(\vec{r})$
(indicated by square bracket ) and is invariant only under a discrete set of translations 
corresponding to lattice vectors $\vec{R}_{i}$.
The functions $h^{(b)}$  and $c^{(b)}$ can therefore be written as a periodic function of 
the center of mass variable $r_{c} = \frac{r_{1}+r_{2}}{2}$  and a continuous function
of the difference variable $\vec{r} = \vec{r}_{2} - \vec{r}_{1}$ .Thus \cite{8}
\begin{eqnarray}
h^{(b)} (\vec{r}_{1},\vec{r}_{2}) = \sum_{G} e^{i\vec{G}.\vec{r}_{c}} \:  h^{(G)} (\vec{r})  \nonumber
\\
c^{(b)} (\vec{r}_{1},\vec{r}_{2}) = \sum_{G} e^{i\vec{G}.\vec{r}_{c}} \: c^{(G)} (\vec{r}).   \label{3}
\end{eqnarray}
Since $h^{(G)}$ and $c^{(G)}$ are real and symmetric with respect to interchange of
$\vec{r}_{1}$ and $\vec{r}_{2}$, they satisfy relations $f^{(G)} (\vec{r}) = f^{(-G)} (\vec{r})$
and $f^{(G)} ( -\vec{r}) = f^{(G)} (\vec{r})$ where $f^{(G)}$ is $h^{(G)}$ or $c^{(G)}$. 

   Substitution of $(2)$ into $(1)$ makes the OZ equation to split into two equations;
one that contains $h^{(0)}, c^{(0)}$ and $\rho$ whereas the other that contains 
$h^{(b)}, c^{(b)}, \rho^{(b)}$ along with $h^{(0)}, c^{(0)}$ and $\rho$,
\begin{align}
h^{(0)} (|\vec{r}_{2} -\vec{r}_{1}|) = c^{(0)} (|\vec{r}_{2} -\vec{r}_{1}|) &+ \rho \int d \vec{r}_{3} c^{(0)}  (|\vec{r}_{3} -\vec{r}_{1}|) \nonumber \\
&\times  h^{(0)} (|\vec{r}_{3} -\vec{r}_{2}|),   \label{4}
\end{align}
\begin{align}
&h^{(b)} (\vec{r}_{1} ,\vec{r}_{2}) = c^{(b)} (\vec{r}_{1} ,\vec{r}_{2})
+  \int d \vec{r}_{3}[ \lbrace \rho^{(b)} (\vr_{3}) c^{(0)} (|\vr_{3} - \vr_{1}|) \nonumber  \\ 
&\: + \rho(\vr_{3}) c^{(b)} (\vec{r}_{1} ,\vec{r}_{3}) \rbrace h^{(0)} (|\vec{r}_{3} -\vec{r}_{2}|)
 + \rho (\vr_{3}) \lbrace c^{(0)} (|\vec{r}_{3} -\vec{r}_{1}|) \nonumber  \\
&\: + c^{(b)} (\vec{r}_{1},{\vr_{3}})  \rbrace  h^{(b)} (\vec{r}_{2},\vec{r}_{3})] .  \label{5}
\end{align}
Eq (4) is the well-known OZ equation of a homogeneous system. It is used along with a closure relation 
to calculate values of $h^{(0)},c^{(0)}$ and their derivatives with
respect to $\rho $ \cite{2,11,12}. Eq (5)is the OZ equation that connects the symmetry 
broken  part of PCFs. We use it to calculate  $h^{(b)} (\vec{r}_{1},\vec{r}_{2})$ from known values
of $ c (\vec{r}_{1},\vec{r}_{2})$ and $\rho (\vec{r})$.

   Using the relation \cite{2,13}
\begin{eqnarray}
\frac{\delta^{n}c(\vec{r}_{1}, \vec{r}_{2})}{\delta \rho (\vec{r}_{3} )\cdots \delta \rho (\vec{r}_{n}) } =
c_{n+2} (\vec{r}_{1},\cdots ,\vec{r}_{n}), \label{6}
\end{eqnarray}
where $c_{m}$  is the $m-$ body direct correlation function, and the functional Taylor expansion,
one can write the following series for $c^{(b)}(\vec{r}_{1},\vec{r}_{2})$ \cite{10},
\begin{align}
&c^{(b)}(\vec{r}_{1},\vec{r}_{2}) = \int d \vec{r}_{3} {c_{3}}^{(0)}(\vec{r}_{1},\vec{r}_{2},\vec{r}_{3}; \rho) (\rho (\vec{r}_{3}) - \rho)    \nonumber \\ 
&\hspace*{-0.5cm} +\frac{1}{2} \int d \vec{r}_{3} \int d \vec{r}_{4} {c_{4}}^{(0)}(\vec{r}_{1},\vec{r}_{2},\vec{r}_{3},\vec{r}_{4} ;\rho)
 (\rho (\vec{r}_{3}) - \rho) (\rho (\vec{r}_{4})-\rho) + \cdots  , \label{7}
\end{align}
where $c^{(0)}_{n}$  is the $n-$ body direct correlation function of a homogeneous system of density 
$\rho$ and $(\rho (\vec{r})-\rho) = \sum_{G} \rho_{G} \exp (i \vec{G}.\vec{r})$. The vales of 
$c_{m}^{(0)}$ are found from exact relations \cite{13},
\begin{eqnarray}
\hspace*{-0.5cm}
\frac{\partial^{n}c^{(0)}(\vec{r}, \rho)} {\partial \rho^{n} } = \int d \vec{r}_{3} \cdots \int d \vec{r}_{n+2}
{c^{(0)}}_{n+2} (\vec{r}_{1},\cdots ,\vec{r}_{n+2}),  \label{8}
\end{eqnarray}

The values of $c^{(0)}(\vec{r}, \rho)$ and  $\frac{\partial^{n}c^{(0)}(\vec{r}, \rho)} {\partial \rho^{n} }$
 are calculated from the OZ	 equation (4) and the Roger-Young closure 
 relation \cite{14} for a system of soft disks 
 interacting via reduced pair potential $\beta u(r)= \frac{\Gamma}{r^{3}} $ where $\beta$ is the inverse 
 temperature in unit of the Boltzmann constant and distance $r$  is measured in unit of 
 $(\frac{1}{\rho})^{\frac{1}{2}} $. The system freezes into a hexagonal crystal at $\Gamma \simeq 10 $ 
 \cite{12,15}. The values of  $\frac{\partial^{n}c^{(0)}(\vec{r}, \rho)} {\partial \rho^{n} }$ 
 and the factorization \textit{ansatz} \cite{11,16} have been used to find values of $c^{(0)}_{n + 2}$ 
 from (8) which are then used in (7) to calculate $c^{(G)} (\vr)$ \cite{11,12}

\begin{figure}[ht]\label{fig-1}
\includegraphics[height=2.5in, width=3.0in,clip]{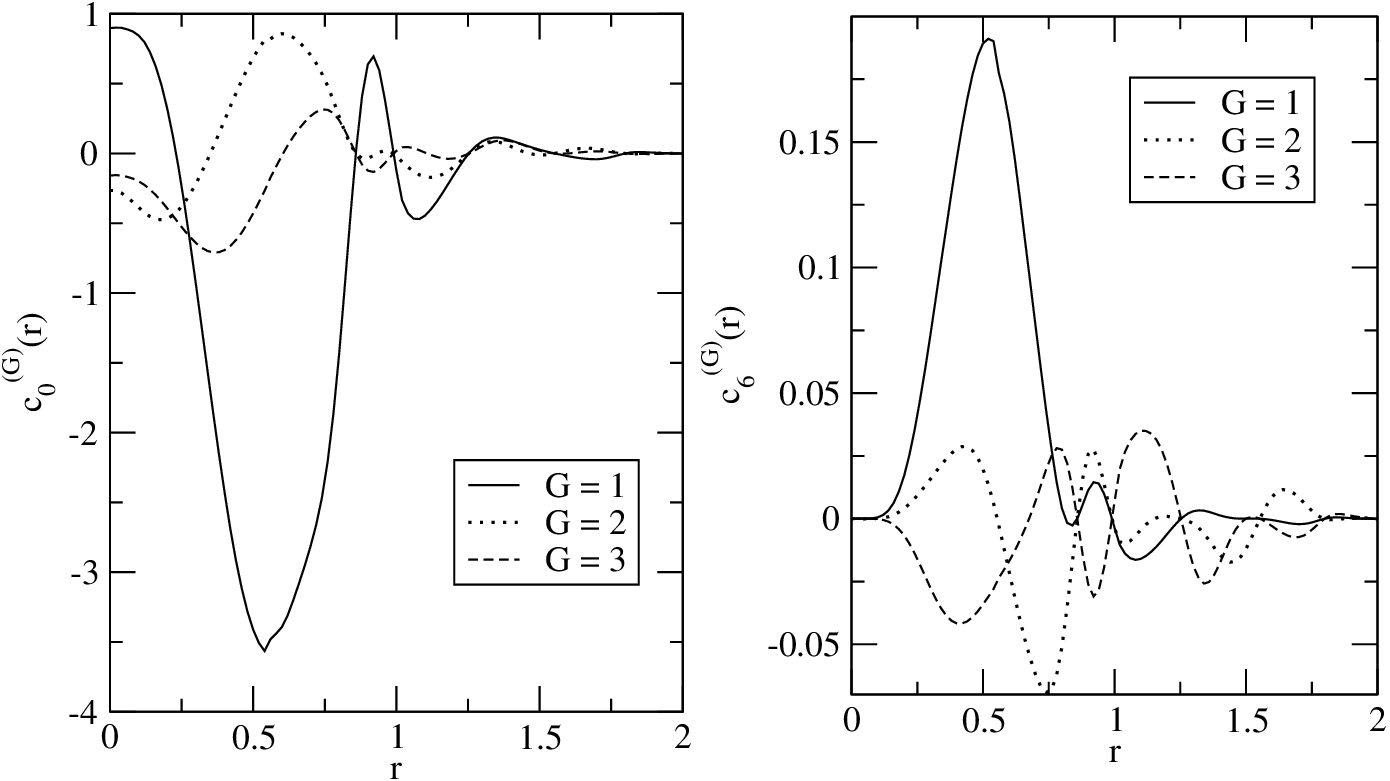}
\caption{Harmonic coefficients  $c_{M}^{(G)} (r)$ for RLV's of first three sets. The full line
represents values for the first set, the dotted line of the second set and dashed line of the 
third set. the distance r is in unit of $(\frac{1}{\rho})^{\frac{1}{2}}$ for a system of soft 
disks interacting via a reduced pair potential  $\beta  u(r) =\Gamma/r^{3} (\Gamma =12 )$  }
\end{figure}
 
  In Fig.(1) we plot $c_{M}^{(G)}(r) $ which for a two- dimensional crystal is defined  as \cite{12}
\begin{eqnarray}
c^{(G)} (\vec{r}) = \sum_M \: (i)^{M} \: c_{M}^{(G)} (\vec{r}) \: e^{iM\phi_{G}} \: e^{-iM\phi_{r}}  \label{9}
\end{eqnarray}
where $M=0, \pm 6$ for a hexagonal lattice ,$\phi_{G}$ and $\phi_{r}$ 
are angles of vectors $\vec{G}$ and $\vec{r}$ with respect to a space fixed 
coordinate frame. The results given in the figure are for $\Gamma = 12$ and found from 
the first term of (7). The values of ${c_{M}}^{(G)}$ is found to depend on values of 
order parameters $\rho_{G}$  and of RLVs. The values plotted are for RLVs of the first three 
sets. We see that for different set of $G$, ${c_{M}}^{(G)}(\vec{r})$ varies with $r$
in different way. The value becomes negligible for $r > 1.5 $ in all cases. 
For any given value of $G$, the value of ${c_{0}}^{(G)}(\vec{r})$ is
about an order of magnitude higher than ${c_{6}}^{(G)}(\vec{r})$ at 
their maxima and minima. As the magnitude of $G$ increases, value of
${c_{M}}^{(G)}{(\vec{r})}$ decreases and after sixth set the value becomes 
negligible. As one goes at lower temperature/higher density more sets may contribute but the 
number will always remain few.

The usefulness of this method, however, depends on the convergence of the series (7). 
It has been shown in refs. \cite{11,12} that at the melting point $c^{(b)}$ is 
accurately approximated by the first term of (7). The following identity  
(see \cite{17,18}), can be used to find contributions made by different terms 
of (7) and therefore its convergence :
\begin{align}
&\nabla_{1} \ln \rho (\vr_{1}) = \int d \vr_{2} c^{(0)} (|\vr_{2} - \vr_{1}|)   
 + \sum_{G}  \int d \vr_{2} \nonumber \\
&\: \times [\nabla_{2} \rho (\vr_{2})+ i (\rho(\vr_{2}) - \rho) \vec{G} ] e^{i\vec{G}\cdot \vr_{c}} 
\tilde{c}^{(G)} (\vr_{2} -\vr_{1}),   \label{10}
\end{align}
where,
$\tilde{c}^{(b)} (\vr) = \int^{1}_{0} d \xi \int^{1}_{0} d \lambda c^{(G)} (\vr,\lambda \rho, \xi \rho_{G})$. 

The l.h.s. of (10) is evaluated using 
$\rho (\vr_{1})=(\alpha/ \pi) \sum_{i} \exp {[- \alpha (\vr -\vec{R_{i}} )^{2}]} $ 
and the r.h.s. $\rho(\vr)=\rho +\sum_{G} \mu_{G} \exp [-i \vec{G} \cdot \vr] $ 
where $\rho_{G} = \rho \exp[-G^{2}/ 4 \alpha]$. The value of $\alpha$ is taken 
100 \cite{12,19}. When values of $c^{(0)} (r)$ and of $c^{(G)} (\vr)$ found from 
the first term of (7) are used, the r.h.s. of (10) 
gives values which are on the average $20 \%$ higher 
than the values found from the l.h.s. This is an estimate of the contribution expected to 
come from higher order terms of (7), which as shown in ref \cite{11} can be 
evaluated using the procedure describe above. We now proceed to calculate 
$h^{(b)} (\vr_{1},\vr_{2})$.
  
 Using (3) we can write (5) as (see \cite{18}) 
\begin{align}
& h^{(G)} (\vec{r}) = \frac{1}{V} \sum_{k} H^{(G)} (\vec{k}) e^{i \vec{k}\cdot \vr} + 
\frac{\rho}{V} \sum_{G_{1}} \sum_{k} \nonumber \\
&\: h^{(G_{1})} \left( \vec{k} - \frac{1}{2} \vec{G} -\frac{1}{2} \vec{G_{1}} \right) 
 \left[  \mu_{G-G_{1}}  c^{(0)} \left(-|\vec{k} + \frac{1}{2} \vec{G}|\right) \right.  \nonumber \\ 
&\:\left. + \sum_{G_{2}} \mu_{G-G_{1} -G_{2}} c^{(G_{2})} \left( \frac{1}{2} \vec{G_{2}} -
\frac{1}{2}\vec{G} -\vec{k} \right) \right] e^{i\vec{k}. \vec{r}} , \label{11}
\end{align}
where 
\begin{align}
&H^{(G)} (\vec{k}) = c^{(G)} (\vec{k}) +\rho \: \mu_{G} \: c^{(0)} \left(|\vec{k} -\frac{1}{2} \vec{G}| \right) h^{(0)} \left( -|\vec{k} +\frac{1}{2} \vec{G}| \right) \nonumber \\
 & \: + \rho \sum_{G_{1}} \mu_{G_{1}-G} \: c^{(G_{1})} \left( \vec{k}- \frac{1}{2} \vec{G} +\frac{1}{2} \vec{G_{1}}  \right) h^{(0)} \left(-|\vec{k} + \frac{1}{2}\vec{G_{1}}| \right).  \label{12}
\end{align}
We adopt iterative method (see \cite{18}) to calculate values of $h^{(G)} (\vr)$ from (11). It is
found that the term that contains $c^{(G)}$ in (11) makes small contribution compared to the term that 
contains $c^{(0)}$, therefore some error in $c^{(G)}$ due to truncation of the series (7) will not 
affect the value of $h^{G}(r)$ in any significant way.

\begin{figure}[ht]\label{fig-2}
\includegraphics[height=2.5in, width=3.0in,clip]{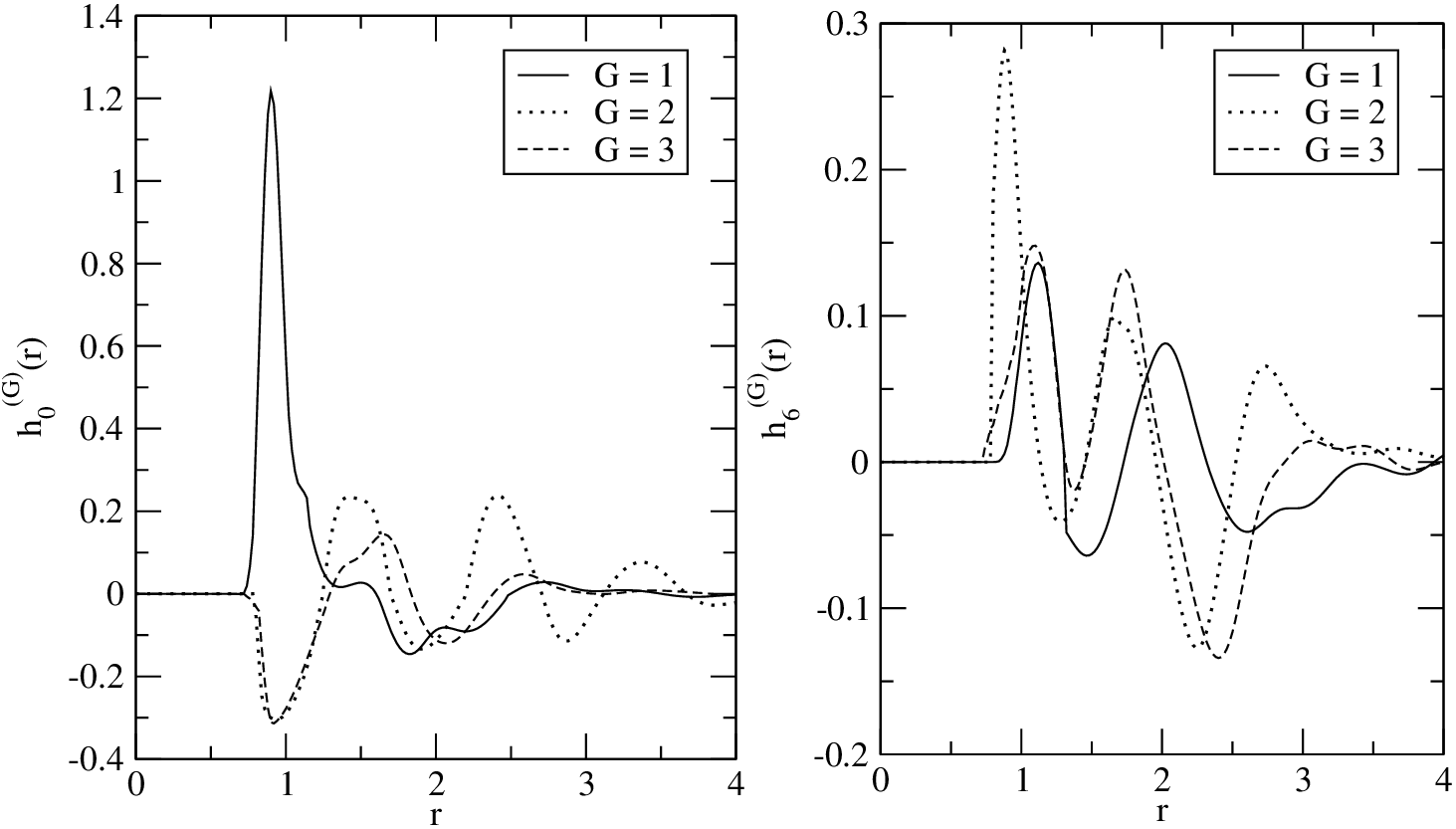}
\caption{Harmonic coefficients $h_{M}^{(G)} (r)$ for RLV's of first three sets.
Notation are same as in Fig. 1}
\end{figure}

 In Fig (2) we plot values of $h^{(G)}_{M} (r)$ defied as   
\begin{eqnarray}
h^{(G)} (\vec{r}) =  \sum_{M} (i)^{M} \: h^{(G)}_{M} (r) \: e^{i M \phi_{G}} \: e^{-i M \phi_{r}},  \label{13}
 \end{eqnarray}
For the first three sets of 
$\vec{G}$ for $\Gamma = 12 $ and $\alpha = 100 $. The values of $h^{(G)}_{M} (r)$ depend on 
values of order parameters $\rho_{G}$ and on values of $\vec{G}$ vectors. For  
$G$ belonging to different sets, values of $h^{(G)}_{M} (r)$ as a 
function of $r$  oscillate about zero in different ways; the maxima and minima 
are located at different values of $r$.
The values decay rapidly and become almost zero for $r > 5$ in all cases.
Unlike $c^{(G)}_{M} (r)$, the values of maxima and minima of $h^{(G)}_{0} (r)$ and 
$h^{(G)}_{6} (r)$ are comparable. As the magnitude of $G$ increases the value of 
$h^{(G)}_{M} (r)$ decreases and as in the case of $c^{(G)}_{M} (r)$ 
after sixth set the values become negligible.

  The accuracy of $h^{(G)} (r)$ can be tested using a Ward identity which in 
this case is represented by the Born-Green -Yoven (YGB) equation \cite{2},
\begin{align}
\nabla_{1}  \ln \rho (\vec{r}_{1}) &=  - \beta \int d \vec{r}_{2} \rho (\vec{r}_{2}) (g^{(0)} (|\vec{r_{2}} -\vec{r}_{1}|) \nonumber \\
 &\qquad + h^{(b)} (\vec{r}_{1} , \vec{r}_{2} )) \nabla_{1} u (|\vec{r}_{2} - \vec{r}_{1}|), \label{14}
\end{align}
where $u (|\vec{r}_{2} - \vec{r}_{1}|)$ is the pair potential. It is solved to give 
\begin{align}
1 = - &\frac{\rho}{2\alpha r_{1}} \sum_{G} i J_{1} (Gr_{1}) \int d \vr \left( 
\frac{\vec{G}}{|\vec{G}|}\cdot \frac{\vr}{|\vr|} \right) \frac{\delta (\beta u(r))}{\delta r} 
\nonumber \\
& \times \left[\mu_{G} g^{(0)} (r) e^{i\vec{G}\cdot \vr} + h^{(G)} (\vr) e^{\frac{1}{2} i \vec{G} \cdot \vr} \right.
\nonumber \\
& \left. + \sum_{G_{1}} \mu_{G-G_{1}} h^{(G_{1})} (\vr) e^{i (\vec{G} -\frac{1}{2}\vec{G_{1}}) \cdot \vr} 
\right]
\label{15}
\end{align}
The values calculated for several values of $r_{1}$ are found to lie between 
$0.92 - 1.10$ indicating that the value of $h^{(G)}_{M} (r) $ 
given in Fig 2 are reasonably accurate.

\begin{figure}[ht]\label{fig-3}
\includegraphics[height=2.5in, width=3.0in,clip]{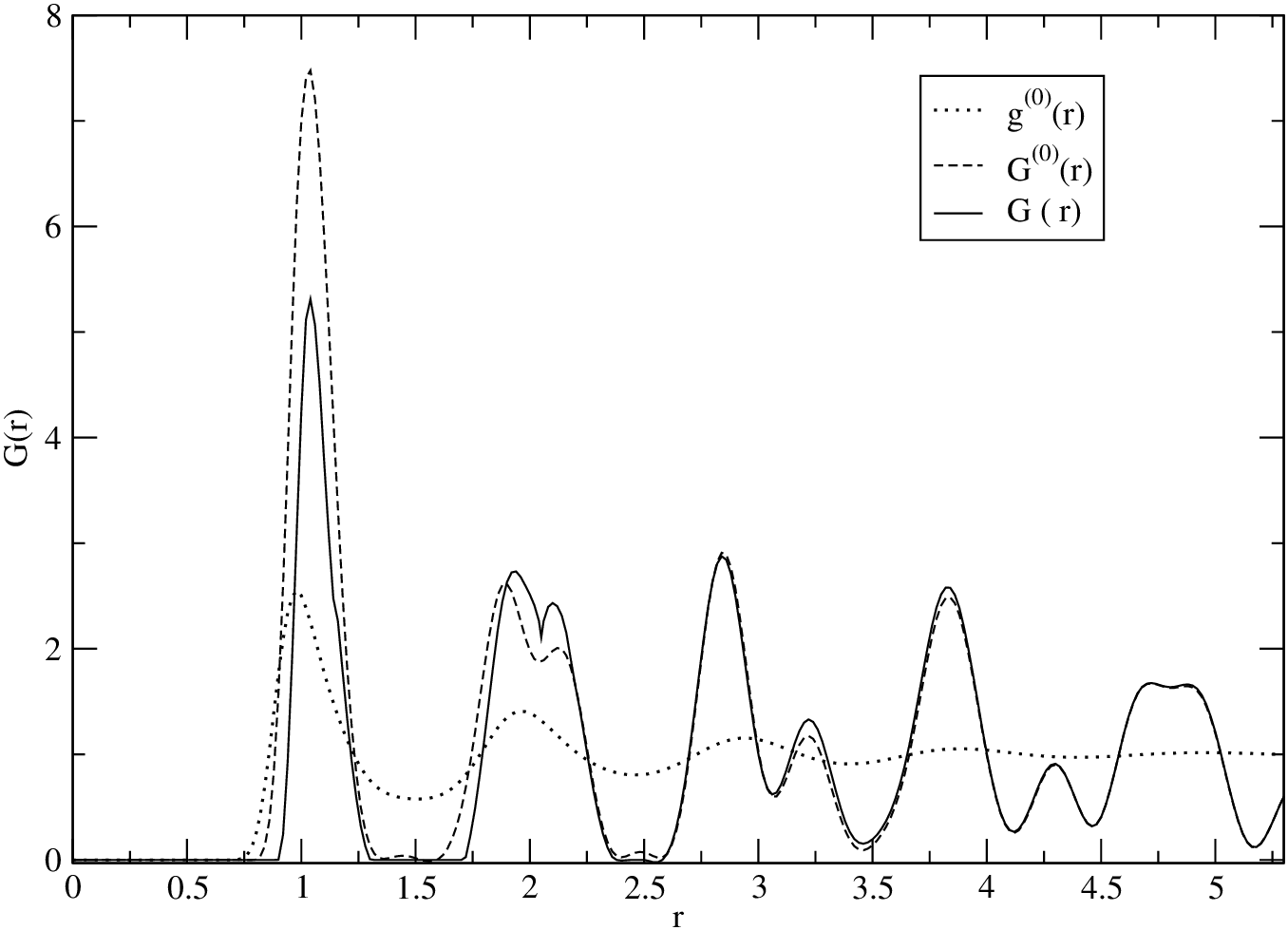}
\caption{ Correlation function $G(r)$ (see Eq. (16)) as a function of {\it{r}}. 
$G^{(0)}$ refers to values found without symmetry breaking part of 
$h(\vr_{1},\vr_{2})$. $g^{(0)} (r) $ is the radial distribution function 
in a fluid of the average density of the crystal. }
\end{figure}

  In computer simulations \cite{20} and in experiments \cite{21}  $g(\vec{r}_{1},\vec{r}_{2})$ 
is reduced to a function of one variable, 
\begin{align}
G(r) & = \frac{1}{\rho^{2} V} \int_{0}^{2\pi} \frac{d \phi_{r}}{2 \pi} \int d \vec{r_{1}} 
\rho (\vec{r}_{1}) \rho (\vec{r_{2}}) \nonumber \\
& \qquad \times \left[g^{(0)} (|\vec{r_{2}} - \vec{r}_{1}|) + h^{(b)} (\vec{r_{1}} , \vec{r}_{2}) \right], 
\label{16} 
\end{align}
where $\phi_{r}$ is the angle of vector $\vr = \vr_{2} - \vr_{1}$. 
In the fluid $G(r)$ reduces to the radial distribution function $g^{(0)} (r)$.
The values of $G(r)$ plotted in Fig (3) for $\Gamma = 12 $ as a function of r
are in good qualitative agreement with the values found from  
simulations and experiments for soft colloidal particles \cite{20,21}; as
the pair potentials in these work are different we can not expect 
quantitative agreement. In the figure we also give the values of $g^{(0)}(r) $ and 
the value of $G^{(0)}(r)$ found without including the contribution of
$h^{(b)} (\vec{r}_{1} ,\vec{r}_{2})$. It is seen that the effect 
of $h^{(b)} (\vec{r}_{1} ,\vec{r}_{2})$ is limited to the first 
two peaks of $G(r)$ only.

   In conclusion; we developed a method to calculate PCFs in a crystal. 
This method is based on separating the correlation functions into the symmetry conserving 
and the symmetry broken parts. The symmetry conserving parts are calculated using the 
IET of homogeneous fluids. The symmetry broken part of $c (\vec{r}_{1}, \vec{r}_{2})$
is calculated from a series written in powers of order parameters. The series involves 
three- and higher-body direct correlation functions of a homogeneous system which are  
calculated from relations connecting them with the density derivatives of 
$c^{(0)} (r,\rho)$ . It is shown that near the melting point most of the contribution 
comes from the first term of the series. As one goes deeper into the solid 
higher order terms may start contributing which can easily be 
calculated using the method used for 
the first term and described in \cite{11}. The term $h^{(b)}(\vec{r}_{1}, \vec{r}_{2}) $ is 
calculated from the OZ	equation using an iterative 
method. The values found for $h^{(G)} (\vr)$ for $G$ of different sets 
differ substantively from each other, but in all cases they decay to  zero 
for $r \ge 5$. This shows that $\rho(\vec{r}_{1}) \rho(\vec{r}_{2}) h(\vec{r}_{1}, \vec{r}_{2})$
is a short range function. The accuracy of the values found for 
$h (\vec{r}_{1} ,\vec{r}_{2})$ has been examined using the
YGB equation that relates $\rho(\vec{r}_{1})$ to the 
integral of $h (\vec{r}_{1} ,\vec{r}_{2})$. Fig 3 shows that it is the first 
two peaks of $G(r)$ which are mainly affected due to $h^{(G)} (\vr)$ , this is  
due to mutual cancellations of values of different $G$ sets. Though the results 
reported here are for a two-dimensional solid, it is straightforward to apply
the theory to three-dimensional crystals. On the basis of these 
results we conclude that the method proposed here can be used to obtain 
accurate and detailed informations about PCFs in crystals.

%------------------------------------------------------------------------

%------------------------------------------------------------------------
%----------------------------------------------------------------------------------
\end{document}

\begin{center}
{\underline{\bf{Supplemental Online  Material}}}
\end{center}
In this supplemental online material we give derivations of some of the equations 
given in the paper and calculational details.
\section*{1. Derivation of Eq.(10)}
In a density functional formalism  one gets a relation connecting $\rho(\vr_{1}) $ with the single 
particle direct correlation function as [2,13]
\begin{eqnarray}
\rho (\vr_{1}) = \frac{e^{\beta (\mu- u^{*}(\vr_{1})) + c(r_{1},[\rho]) } }{\wedge}  \label{S.1}
\end{eqnarray}
where $\mu$ is the chemical potential, $u^{*} (\vr_{1})$ potential at $\vr_{1}$ due to 
external field, $\wedge$ is the cube of the thermal wavelength associated with 
a particle and $c^{(1)}(\vr)$ is the one-body direct correlation function. The functional 
derivative of $c^{(1)}$ with respect to $\rho (\vr)$ defines the direct pair correlation 
function (DPCF),
\begin{eqnarray}
\frac{\delta c^{(1)} (\vr_{1}, [\rho ]) }{\delta \rho (\vr_{2}) } = c (\vr_{1}, \vr_{2}, [\rho])
, \label{S.2}
\end{eqnarray} 
In a crystal the DPCF is written as a sum of symmetry conserving and symmetry broken terms. Thus
\begin{align}
\frac{\delta c^{(1,0)} (\vr, [\rho ]) }{\delta \rho (\vr_{2}) }& = c (|\vr_{2} - \vr_{1}|, \rho) 
\label{S.3}
\end{align}
and 
\begin{align}
\frac{\delta c^{(1,b)} (\vr, [\rho ]) }{\delta \rho (\vr_{2}) }& = c^{(b)} 
(\vr_{1}, \vr_{2}, [\rho]) , \label{S.4}
\end{align}
where the superscripts $(0)$ and $(b)$ refer to symmetry conserving and symmetry broken parts
of direct correlation functions. The values of $c^{(1,0)}$ and $c^{(1,b)}$ are found from functional 
integrations of Eqs (\ref{S.3}) and (\ref{S.4}), respectively. In this integration the system
is taken from some initial density to final density along a path in the density space; the result is 
independent of the path of integration. As the symmetry conserving part $c^{(0)}$ is a function of
density only, the integration in Eq.(\ref{S.3}) is done taking an isotropic system of density
$\rho$ as a reference. This gives,
\begin{eqnarray}
c^{(1,0)} (r_{1},[\rho]) = c^{(1)} (\rho) +  \int d \vr_{2} (\rho(\vr_{2}) - \rho) c^{(0)} 
(|\vr_{2}-\vr_{1}|,\rho).  \label{S.5} 
\end{eqnarray}
where $c^{(1)} (\rho)$ is a density dependent constant. In order to do functional integration of (\ref{S.4}) 
in which $c^{(b)}[\rho] $ depends on the order parameter in addition to the averaged density, 
we chose a path in density space  which is characterised by two 
parameters $\lambda $ and $\xi$ [11,12]. These parameters vary from 0 to 1.
The parameter $\lambda$ raises the average density from 0 to the final value $\rho $ 
on its variation from 0 to 1, whereas the parameter $\xi $ raises the 
order parameters from 0 to their final value $\rho_{G}$ for each $G$ 
when it varies from 0 to 1. The integration gives,
\begin{align}
c^{(1,b)} (r,[\rho]) =  \int d \vr_{2} (\rho(\vr_{2}) - \rho) \tilde{c}^{(b)} 
(\vr_{1},\vr_{2},[\rho]) , \label{S.6} 
\end{align}
where 
\begin{align}
\tilde{c}^{(b)} (\vr_{1},\vr_{2},[\rho]) = \int_{0}^{1} d \xi \int_{0}^{1} d \lambda 
c^{(b)} (\vr_{1},\vr_{2},\lambda \rho, \xi \rho_{b}) .  \label{S.7}
\end{align}
While integrating over $\lambda$ the order parameters $\rho_{G}$ are kept fixed, and while integrating over 
$\xi$ density is kept fixed. The result does not depend on the order of integration [9,10]. 
From above equations we get after taking $ u^{*}(\vec{r}) = 0 $,
\begin{align}
\ln \rho(\vr_{1})& = \ln \rho + \int d \vr_{2} (\rho (\vr_{2}) - \rho) c^{(0)} (|\vr_{2} - \vr_{1}|,\rho) \nonumber \\
 & \quad +  \int d \vr_{2} (\rho (\vr_{2}) - \rho) \tilde{c}^{(b)} (\vr_{1},\vr_{2},[\rho]), \label{S.8}
\end{align}  
where 
\begin{align}
\ln \rho = \beta \mu + c^{(1)} (\rho) - \ln \wedge . \nonumber
\end{align}
Giving spacial displacement $\vec{\delta}$ to position vectors and taking the limit 
$\delta \rightarrow 0 $ one gets Eq (10)) :
\begin{align}
\nabla_{1} \ln \rho (\vr_{1}) & = \int d \vr_{2}  \nabla_{2} \rho (\vr_{1}) 
c^{(0)} (|\vr_{2} - \vr_{1}|, \rho)  \nonumber \\
& \qquad + \sum_{G_{1}} \int d \vr_{2} [\nabla_{2} \rho (\vr) + i (\rho (\vr_{2}) - \rho) \vec{G_{1}}] 
e^{i\vec{G_{1}} \cdot \vr_{c}}  \tilde{c}^{(\vec{G}_{1})}  (\vr_{2} - \vr_{1}) ,  \label{S.9}
\end{align}
where  $ \tilde{c}^{(b)} (\vr_{1},\vr_{2}) = \sum_{G_{1}} e^{i\vec{G_{1}}\cdot \vr_{c}} 
c^{(G_{1})} (\vr_{2} -\vr_{1})   $.

\section*{2. Evaluation of $h^{(b)} (\vr_{1},\vr_{2})$ }
The OZ equation (5) can be written as
\begin{eqnarray}
h^{(b)} (\vr_{1},\vr_{2}) = H^{(b)} (\vr_{1},\vr_{2}) + \int d \vr_{3} h^{(b)} (\vr_{1},\vr_{3}) \rho (\vr_{3})
\left[ c^{(0)} (|\vr_{3} - \vr_{2}| ) + c^{(b)} (\vr_{2},\vr_{3})\right]             \label{S.10}
\end{eqnarray}
where 
\begin{align}
H^{(b)} (\vr_{1},\vr_{2})& = c^{(b)} (\vr_{1},\vr_{2}) + \int d \vr_{3} h^{(0)} (|\vr_{3}- \vr_{1}|)  \nonumber \\
 & \quad [ \rho^{(b)} (\vr_{3}) c^{(0)} (|\vr_{3} - \vr_{2}|) +\rho (\vr_{3}) c^{(b)} (\vr_{2},\vr_{3})]             \label{S.11}
\end{align}
\subsection{Derivation of Eq (11)}
Consider a term of Eq (\ref{S.10})
\begin{align}
I  = \int d \vr_{3} h^{(b)} (\vr_{1},\vr_{3}) \rho (\vr_{3}) c^{(0)} (|\vr_{3} - \vr_{2}|).   \label{S.12}
\end{align}
Substituting 
\begin{align}
h^{(b)} (\vr_{1},\vr_{3}) = \sum_{G_{1}} e^{i \vec{G}_{1} \cdot \left( \frac{\vr_{1}+\vr_{3}}{2}\right)} h^{(G_{1})} (\vr_{3} -\vr_{1}) ,  \nonumber
\end{align}
and $\rho (\vr_{3}) = \rho \sum_{G_{2}} \mu_{G_{2}} e^{i G_{2} \cdot \vr_{3}} $,
where $\mu_{0} = 1$ in  (S.12) we get 
\begin{align}
I   & =\rho \sum_{G_{1}} \sum_{G_{2}} e^{ \frac{1}{2}i \vec{G_{1}} \cdot \vr_{1}} \mu_{G_{2}} \int d \vr_{3} 
h^{(G_{1})} (\vr_{3} -\vr_{1}) e^{i (\frac{1}{2} \vec{G_{1}} +\vec{G_{2}}) \cdot \vr_{3} } 
c^{(0)} (|\vr_{3} -\vr_{2}|)   , \nonumber \\
 & = \rho \sum_{G_{1}} \sum_{G_{2}} \mu_{G_{2}} e^{i \frac{1}{2} \vec{G_{1}} \cdot \vr_{1}}
\frac{1}{V^{2}} \sum_{q_{1}} \sum_{q_{2}} h^{G_{1}} (\vec{q_{1}}) c^{(0)} (q_{2}) 
e^{-i \vec{q_{1}} \cdot \vr_{1}}  e^{-i \vec{q_{2}} \cdot \vr_{2}}  \nonumber \\
\qquad & \qquad \int d \vr_{3} e^{i \left( \frac{1}{2} \vec{G_{1}} + \vec{G_{2}} +\vec{q_{1}} +\vec{q_{2}} \right)\cdot \vr_{3}} , \nonumber \\
 & = \rho \sum_{G_{1}} \sum_{G_{2}} \mu _{G_{2}} e^{\frac{1}{2} i \vec{G_{1}} \cdot \vr_{1}} \frac{1}{V} 
\sum_{q_{1}} h^{(G_{1})} (\vec{q_{1}}) c^{(0)} (-|\vec{q_{1}} +\frac{1}{2} \vec{G_{1}} +\vec{G_{2}}|) 
e^{-i \vec{q_{1}}\cdot \vr_{1}}  e^{i \left(\vec{q_{1}}+ \frac{1}{2} \vec{G_{1}} + \vec{G_{2}} \right)\cdot \vr_{2} } , \nonumber \\
 & = \rho \sum_{G_{1}} \sum_{G_{2}} \mu_{G_{2}} e^{\frac{1}{2} i(\vec{G}_{1}+\vec{G}_{2})\cdot
(\vr_{1} +\vr_{2})} \frac{1}{V} \sum_{q_{1}} h^{G_{1}} (\vec{q_{1}}) c^{(0)} \left( -|\vec{q_{1}} +\frac{1}{2} \vec{G_{1}} + \vec{G_{2}}| \right) e^{i \left(\vec{q_{1}} +\frac{1}{2} \vec{G_{2}}\right)\cdot \vr } .
 \label{S.13}
\end{align}
Substituting  $\vec{q_{1}} + \frac{1}{2}\vec{G_{2}} =\vec{k} $  and  $\vec{G_{1}}+\vec{G_{2}}= \vec{G} $ in
(\ref{S.13}) we get 
\begin{eqnarray}
I = \frac{\rho}{V} \sum_{G}\sum_{G_{1}} \mu_{G-G_{1}} e^{i \vec{G} \cdot \vr_{c}} \sum_{k} h^{(G_{1})} 
\left( \vec{k} - \frac{1}{2}\vec{G} -\frac{1}{2}\vec{G_{1}} \right) 
c^{(0)} \left(-| \vec{k} + \frac{1}{2}\vec{G} |\right)  \label{S.14}
\end{eqnarray}
which is first term in the square bracket of Eq (11). Similarly 
other terms of Eq (11) can be derived from Eqs (\ref{S.10}) and (\ref{S.11})

\subsection{Evaluation of $ h^{(G)}(\vr)$}

Eq(11) is solved to give
\begin{eqnarray}
h^{(G)}_{M} (r) = H^{(G)}_{M}(r) + \sum_{n=1}^{2} h^{(G,n)}_{M} (r),    \label{S.15}
\end{eqnarray}
where
\begin{align}
H^{(G)}_{M} (r) &= c^{(G)}_{M} (r) + \rho \mu_{G} \sum_{m} (-1)^{m} J_{M-m} \left( \frac{1}{2} Gr \right)
B^{(1)}_{m} (r,G)   \nonumber \\
 & \qquad + \rho {\sum_{G_{1}}} ^{'} \mu_{G_{2}} \delta_{\vec{G},\vec{G_{1}} + \vec{G_{2}}} \sum_{m}
\sum_{m_{1}} e^{i m (\phi_{G_{1}} - \phi_{G})} e^{i(M-m-m_{1})(\phi_{G_{2}} - \phi_{G})}  
 B_{M,m,m_{1}}^{(2)} (G,G_{1},r)    ,       \label{S.16}  \\
h^{(G,1)}_{M} (r)& = \rho \sum_{G} {\sum_{G_{2}}}^{'} \mu_{G_{2}} \delta_{\vec{G},\vec{G_{1}} + \vec{G_{2}}} \sum_{M_{1}} \sum_{m_{1}} e^{i M_{1}(\phi_{G_{1}} - \phi_{G})} e^{i m_{1}(\phi_{G_{2}} - \phi_{G})} B_{M,m,m_{1}}^{(3)} (G,G_{1},r)    \label{S.17}
\end{align} 
and
\begin{align}
h^{(G,2)}_{M} (r)& = \rho \sum_{G} \sum_{G_{2}} {\sum_{G_{3}}}^{'} \mu_{G_{3}} \delta_{\vec{G},\vec{G_{1}}
+ \vec{G_{2}} + \vec{G_{3}}}  \sum_{M_{1}} \sum_{M_{2}}  \sum_{m_{1}} \sum_{m_{2}} \sum_{m_{3}} \sum_{m_{4}}  
 \nonumber \\ 
 & \qquad \delta_{M,M_{1}+M_{2}+m_{1}+m_{2}+m_{3}+m_{4}} e^{i (M_{1}+m_{1})(\phi_{G_{1}} - \phi_{G})}
 e^{i (M_{2}+m_{3})(\phi_{G_{2}} - \phi_{G})}  \nonumber \\
 & \qquad e^{i (m_{2} + m_{4})(\phi_{G_{3}} - \phi_{G})}
 B_{M,M_{1},M_{2},m_{1},m_{2},m_{3},m_{4}}^{(4)} (G,G_{1},G_{2},r),   \label{S.18}  
\end{align} 
Here
\begin{align}
B^{(1)}_{m} (r,G)& = \int d \vr_{3} c^{(0)} (r_{13}) J_{m} (Gr_{13}) h^{(0)} (r_{23}) 
e^{-im (\phi_{13}- \phi_{12} )}    \nonumber \\
B_{M,m,m_{1}}^{(2)} (G,G_{1},r) &= \int d \vr_{3} c_{m}^{(G_{1})} h^{(0)} (r_{23}) J_{m_{1}} 
(\frac{1}{2} G r_{23}) J_{M-m-m_{1}} (\frac{1}{2} G_{2} r_{13})  \nonumber \\
 & \qquad   e^{-i(M-m_{1})(\phi_{13} - \phi_{12})}
e^{-im_{1}(\phi_{23} -\phi_{12})} \nonumber \\
 B_{M,m,m_{1}}^{(3)} (G,G_{1},r) & =  \int d \vr_{3} h^{(G_{1})}_{M_{1}} (r_{13}) c^{(0)} (r_{23})
 J_{M-M_{1}-m_{2}} \left(\frac{1}{2} Gr_{23} \right)  
  J_{m_{2}} \left(\frac{1}{2} G_{2}r_{13} \right) \nonumber \\
   & \qquad e^{-i(M-M_{1}-m_{2})(\phi_{23} -\phi_{12})} e^{-i(M_{1}+m_{2})(\phi_{13} -\phi_{12})} \nonumber
\end{align}
\begin{align}
B_{M,M_{1},M_{2},m_{1},m_{2},m_{3},m_{4}}^{(4)} (G,G_{1},G_{2},r)& =\int d \vr_{3} h^{(G_{1})}_{M_{1}} (r_{13})
 c^{(G_{2})}_{M_{2}} (r_{23}) J_{m_{1}} \left(\frac{1}{2}G_{1} r_{23} \right)  \nonumber \\
&  J_{m_{2}} \left(\frac{1}{2}G_{3} r_{23} \right)  J_{m_{3}} \left(\frac{1}{2}G_{2} r_{13} \right) 
 J_{m_{4}} \left(\frac{1}{2}G_{3} r_{13} \right)  \nonumber \\
 & e^{-i(M_{2}+m_{1}+m_{2})(\phi_{23} -\phi_{12})} e^{-i(M_{1}+m_{3}+m_{4})(\phi_{13} -\phi_{12})}   \nonumber 
\end{align}
$\delta_{\vec{G},\vec{G_{1}} +\vec{G_{2}}}$ indicates $\vec{G}= \vec{G_{1}} + \vec{G_{2}}$ and prime on summation in Eqs. (\ref{S.16})-(\ref{S.18}) indicates that for $\vec{G_{2}} = 0$, $\mu _{G_{2}} =1 $ 
and $m_{1} = M - m$. In above equations $J_{m}(x) $ are the Bessel functions of the first kind of 
integral order. 

 We first calculated $H^{G}_{m} (r)$ as it involves known quantities $c^{(0)} (r), c^{(G)}_{m} (r)$ and
$h^{(0)} (r)$ and prepared initial guess for $h^{(G)}_{m} (r)$ by reducing the value by a factor of 10. 
Since $h^{G}_{m} (r)$ is expected to be zero inside the core we have used this fact in successive 
iterations. The mixing procedure in which a small percentage  $ < 10  \%$ of the output is mixed 
with the earlier input to prepare a new input has been used. After 20 such iterations we took the 
output as the new initial guess and repeated the same procedure for five times. The result was 
taken as a new input and iteration were repeated till we got the accuracy of the order of $10^{-3}$.
The final result for $\Gamma = 12 $ is plotted in Fig 2.